\documentclass[11pt,a4paper]{article}
\usepackage{jcappub}
\usepackage{array,dcolumn}
\usepackage{calc,tabularx,mathrsfs}

\usepackage{verbatim,enumerate}
\usepackage{multirow}
\usepackage{setspace}
\usepackage{physics}
\usepackage[toc,page]{appendix}
\usepackage{xspace}
\usepackage{slashed}
\allowdisplaybreaks[1]
\newlength{\figurewidth}
\newcommand{\beq}{\begin{equation}}
\newcommand{\eeq}{\end{equation}}
\newcommand{\bea}{\begin{eqnarray}}
\newcommand{\eea}{\end{eqnarray}}
\newcommand{\ba}{\begin{array}}
\newcommand{\ea}{\end{array}}

\newcommand{\pt}{\partial}

\newcommand{\al}{\alpha}
\newcommand{\bt}{\beta}
\newcommand{\g}{\gamma}
\newcommand{\ep}{\epsilon}
\newcommand{\ta}{\theta}
\newcommand{\lam}{\lambda}

\newcommand{\G}{\Gamma}
\newcommand{\nb}{\nabla}
\newcommand{\de}{\delta}

\newcommand{\OM}{\Omega}

\newcommand{\sg}{\sigma}
\newcommand{\Sg}{\Sigma}
\newcommand{\kp}{\kappa}

\title{Non-locality effect on the entanglement entropy in deSitter}
\author[a]{Gaurav Narain,}
\author[a,b]{Hai-Qing Zhang}
\affiliation[a]{
Centre for Gravitational Physics, Department of Space Science, \\
Beihang University, Beijing 100191, China.}
\affiliation[a,b]{
International Research Institute for Multidisciplinary Science, \\
Beihang University, Beijing 100191, China
}
\emailAdd{gaunarain@gmail.com}
\emailAdd{hqzhang@buaa.edu.cn}
\abstract{
We investigate the effect of infrared non-locality on the entanglement between two 
causally separated open-charts in the deSitter space-time. 
Inspired by the work of Maldacena and Pimentel who gave a precise 
methodology for the computation of the long-range entanglement 
for the local massive scalar field theory on deSitter space-time, we aim 
to investigate the change in behaviour of this long-range entanglement 
due to the presence of infrared non-locality in the theory. 
By considering a non-local scalar field theory where the non-locality 
becomes important in the infrared, we follow the footsteps of 
Maldacena and Pimentel to compute the entanglement entropy of the 
free non-local scalar field theory in the Bunch-Davies vacuum. 
It is found that the presence of infrared non-locality 
will have strong effect on the long-range entanglement. In some 
case it is noted that if the strength of non-locality is large then it 
will tend to decrease the long-range entanglement in the infrared.
We also consider the behaviour of R\'enyi entropy, where a 
strong role of non-locality on the entropy is noticed.
}

\arxivnumber{1812.08667}

\begin{document}
\maketitle
%
%

\section{Introduction}
\label{intro}

Entanglement entropy offers a very valuable insight in to the study of 
long-range quantum phenomenon which occurs in condensed matter system
(see \cite{Amico:2007ag,Horodecki:2009zz} and the references therein). 
This powerful concept also allows one to investigate the 
nature of long-range correlations in the field theories, as it gives  
a way to understand the behaviour of correlation 
between different localised subsystems of a quantum system. 
Local quantum field theory is a well established framework
where meaningful computation can be performed. Moreover, 
in local QFTs which are ultraviolet finite it is known that the 
entanglement entropy follows area law 
\cite{Bombelli:1986rw, Srednicki:1993im}. However, 
not much is known about the situation in the case when
non-locality is present in the system. Non-locality is indeed 
intriguing in the sense the effect it can have on the long-range entanglement. 
Recently in \cite{Narain:2018nfh} a first attempt has been made to 
investigate the effect of infrared non-locality on the 
behaviour of entanglement entropy in anti-deSitter space-time.

Non-locality is a very crucial feature of quantum field theories 
which arises naturally at low-energies due to the running of 
couplings because of quantum corrections. For example 
in quantum theory if the coupling has energy dependence 
$g(\mu)$ (where $\mu$ is the running energy scale), then 
such running can be incorporated in the action 
by generalising the coupling $g(\mu) \to g(-\Box)$, 
where $\Box$ is the square of covariant
derivative. This offers a non-local infrared modifications of the 
field theory. In fact such kind of non-local modification 
has been extensively studied in \cite{Maggiore:2016fbn},
with the aim of obtaining an infrared non-local modification to gravity.
Such infrared modification of gravity has recently been a 
subject of great interest due to its ability in generating a
late time cosmic acceleration mimicking dark energy  
\cite{Maggiore:2016gpx,Narain:2017twx,Narain:2018hxw}.
Universe undergoing an accelerated expansion can be 
beautifully represented as a deSitter space-time, thereby 
putting a strong emphasis on the importance of studying the 
field theory on deSitter space-time. 

Recently, studies on non-local theory have gained momentum 
in regard to ultraviolet modification of various QFTs where it is 
seen that such non-local modifications offer a well-behaved 
UV finite theory which can be super-renormalizable 
\cite{Moffat:2010bh,Biswas:2010zk,Modesto:2011kw,Modesto:2014lga,Modesto:2017sdr}. 
Such UV modification of theory by an addition of non-locality 
provides an extra suppression factor in the propagator of theories 
at high-energies thereby rendering the theory well-behaved in UV.
Non-local theories have also been investigated in relation to black-hole information 
loss paradox \cite{Kajuri:2017jmy,Kajuri:2018myh}. 
Our aim in this paper is to investigate the infrared low-energy effects 
of non-local modifications of theory. We wish to do this 
by investigating the entanglement between the two causally 
dis-connected charts of deSitter space-time. 
In cosmological scenarios we expect that entanglement 
could exists beyond the Hubble horizon as deSitter expansion 
will eventually rips off a pair of particle which were created within 
the causally connected region (inside Hubble radius). If the system 
has non-locality (which is generated by some mechanism) and it 
becomes dominant in infrared, then such 
non-locality will very-likely affect the entanglement of the two 
causally disconnected regions of deSitter space. It is our aim 
to study this effect of non-locality on the entanglement between 
two causally disconnected regions of deSitter.

In the paper we consider a non-local scalar field theory which resides on 
a fixed space-time (deSitter space-time for this paper). 
Here the non-locality arises once one of the field is 
integrated out from the coupled system of local theory. The residual action
describes the behaviour of the massive non-local 
scalar field. This non-local theory has recently been investigated 
in various contexts \cite{Kajuri:2018wow,Narain:2018rif,Narain:2018nfh}.
The purpose is to understand the effect of such infrared non-locality 
on the low-energy physics. Interestingly it is seen in \cite{Narain:2018rif} 
(see the references therein for the IR problem of the dS propagator)
that the presence of the non-locality offers a well-behaved infrared 
finite deSitter propagator of the scalar field theory which decays 
at large time scale. DeSitter propagator of massless local scalar field 
has been a long standing problem where the propagator gets 
IR divergences and the massless limit of massive propagator is not 
well-defined. It was found in \cite{Narain:2018rif} that the presence of 
non-locality helps in overcoming these issues very elegantly thereby 
advocating the importance of non-local effects in the IR physics.
Motivated by this we then investigated the effect of such non-locality 
on the behaviour of entanglement entropy on AdS \cite{Narain:2018nfh}
where the universal finite part was seen to have a characteristic 
oscillating feature due to the presence of non-locality. 

These studies pushed us further to explore the effect this 
infrared non-locality will have on the entanglement between 
two causally disconnected regions of deSitter space-time
which is the work presented in this paper.
In \cite{Maldacena:2012xp} the authors investigated the 
nature of the long-range entanglement entropy on deSitter 
space-time, which they found it to be absent in flat space-time 
and is only specific to deSitter space-time. They described an elegant 
way to specifically extract the part of entanglement entropy 
which captures information about the long-range entanglement. 
We follow the same footsteps described in \cite{Maldacena:2012xp}
to investigate whether the non-locality can play a significant role 
in affecting the long-range entanglement. As scalar fields play an 
important role in cosmology in understanding both early and late-time universe, 
it is therefore important to investigate the effect non-localities 
(which could have arisen either because of quantum effects 
or due to some other mechanism) will have on the long-range entanglement. 

The paper is organised as follows. In section \ref{NLsca} we 
consider the toy model for the non-local theory. In section 
\ref{eeDS} we give a setup of the computation of the 
entanglement entropy in dS. in section \ref{modeCom}
we compute the modes of the non-local scalar field 
on deSitter. In section \ref{denmat} we compute the 
density matrix for our theory and correspondingly 
compute the entanglement entropy. We compute the 
R\'enyi entropy and entanglement negativity for our 
non-local system in section \ref{RenNeg}. We conclude the 
paper with a summary of results and discussion in 
section \ref{conc}.

\section{Non-local scalar field}
\label{NLsca}

In this section we consider a scalar field theory which leads to 
non-local action when one of the scalar gets decoupled from system. 
Consider the following action,
\beq
\label{eq:slocal}
S = \int {\rm d}x \sqrt{-g} \biggl[
\frac{1}{2} (\pt \phi)^2 + \frac{1}{2} (\pt \chi)^2 + \frac{m^2}{2} \phi^2 
- \lam \phi \chi
\biggr] \, ,
\eeq
where $\phi$ and $\chi$ are two scalar fields on curved non-dynamical 
background, $\lam$ is their interaction strength and $m$ is mass of scalar $\phi$. 
The equation of motion of two fields give $(-\Box + m^2)\phi - \lam \chi=0$ 
and $-\Box \chi - \lam \phi =0$. Integrating out $\chi$ from the second equation 
of motion yields $\chi = \lam (-\Box)^{-1} \phi$. This when plugged back into 
the action (\ref{eq:slocal}) yields a massive non-local theory for scalar $\phi$
whose action is given by,
\beq
\label{eq:NLact}
S_{NL} = \int {\rm d}x \sqrt{-g}
\biggl[
\frac{1}{2} (\pt \phi)^2 + \frac{m^2}{2} \phi^2 
-\frac{\lam^2}{2} \phi \frac{1}{-\Box} \phi
\biggr] \, .
\eeq
This non-local action has issues of tachyon. In simple case of 
flat space-time it is noticed that the non-local piece in action reduces 
to $\phi(-p)(-\lam^2/p^2) \phi(p)$ (where $\phi(p)$ is the fourier 
transform of field $\phi(x)$). This piece correspond to something like 
tachyonic mass indicating instability of vacuum. Also, in massless case if we write 
\beq
\label{eq:DeCTr}
\phi_{-} = (\phi+\chi)/\sqrt{2}\, , 
\hspace{5mm} 
\phi_{+} = (\phi-\chi)/\sqrt{2} \, ,
\eeq
then the local action in eq. (\ref{eq:slocal}) can be reduced 
to action of two decoupled free scalar fields 
$\phi_{-}$ and $\phi_{+}$ of mass-square $-\lam$ and $\lam$ respectively. 
\beq
\label{eq:langTach}
S_{\rm FT} = \int {\rm d}x \sqrt{-g} \biggl[
\frac{1}{2} (\pt \phi_{-})^2 + \frac{1}{2} (\pt \phi_{+})^2 
- \frac{\lam}{2} \phi_{-}^2 + \frac{\lam}{2} \phi_{+}^2 
\biggr] \, . 
\eeq
Immediately we notice the field $\phi_{-}$ is tachyonic.
If we change the sign of $\lam$ then $\phi_{+}$ becomes 
tachyonic. This tachyonic issue only disappears for case 
when $\lam=0$ which correspond to two uncoupled 
massless scalar fields. In the massive case things are just more involved, but tachyonic 
issue remains. To see this more clearly we make a linear transformation by writing the 
original set of field ($\phi$ and $\chi$) in terms of new fields ($\phi_1$ and 
$\phi_2$) as 
\beq
\label{eq:transFTphi}
\phi = A_1 \phi_1 + A_2 \phi_2 \, ,
\hspace{5mm}
\chi = B_1 \phi_1 + B_2 \phi_2 \, .
\eeq
By demanding that the transformed Lagrangian is a decoupled system of two scalars 
we get the constraints
\beq
\label{eq:constFTAB}
A_1 A_2 + B_1 B_2 =0 \, ,
\hspace{5mm}
m^2 A_1 A_2 - \lam A_1 B_2 - \lam A_2 B_1 = 0 \, .
\eeq
By expressing $A_1 = \kp B_1$ and $B_2 = - \kp A_2$, one can express the 
transformed lagrangian in terms of $\kp$, $A_2$ and $B_1$. The parameters 
$B_1$ and $A_2$ can be determined by requiring to be nonzero and the kinetic terms 
for the transformed fields to be in canonical form. This gives a 
condition on the parameter $\kp$ to be 
\beq
\label{eq:kpConst}
\lam \kp^2 + m^2 \kp - \lam =0 \, ,
\eeq
and the transformed lagrangian to be 
\beq
\label{eq:TransLagFT}
 S_{\rm FT} = \int {\rm d}x \sqrt{-g} \biggl[
 \frac{1}{2} (\pt \phi_1)^2 + \frac{1}{2} (\pt\phi_2)^2
 - \frac{1}{2} \lam \kp \phi_1^2 + \frac{\lam}{2\kp} \phi_2^2 
\biggr] \, ,
\eeq
where $\kp$ is determined from the equation (\ref{eq:kpConst}). If 
$\kp_1$ and $\kp_2$ are two roots of the eq. (\ref{eq:kpConst}), then 
they satisfy the relation $\kp_1 \kp_2 = -1$. This will mean that 
one of the transformed field will be tachyonic in nature. So the 
tachyonic issue is inherent to the kind of non-local system considered here. 

Such tachyonic problem can be avoided if we consider
$\lam^2 \to -\lam^2$. In this situations the poles of the theory are 
however complex in nature. In particular, when one considers massless theory
where transformation as in eq. (\ref{eq:DeCTr}) can be applied, it 
is found that decoupled action for two scalar fields 
$\phi_{-}$ and $\phi_{+}$ have mass-square $-i\lam$ and $i\lam$ respectively. 
This results in complex mass-poles. 
Field theory with complex poles have been a
well studied subject since the $60$'s 
\cite{Cutkosky:1969fq,Lee:1969fy,Yamamoto:1970gw,Nagy:1970dx},
where unitarity, causality of such theories have been investigated. 
In light of those works one can therefore safely have trust 
in the exploration of such non-local theories and 
consider them as an effective infrared theory which has 
a one particular kind of non-locality. 

\section{Setup for Entanglement Entropy on dS}
\label{eeDS}

In this section we study the entanglement entropy for the above 
non-local quantum field theory on deSitter space-time. We choose 
the standard vacuum-state \cite{Chernikov:1968zm,Bunch:1978yq,Hartle:1983ai}
(which is commonly referred to as Euclidean/Bunch-Davies vacuum)
for the field theory described on non-dynamical background deSitter space-time.

As described in \cite{Maldacena:2012xp}, we consider a spherical surface 
which divides the dS space-time into interior and exterior regions. The entanglement 
between the modes lying on the two sides of the surface is computed
by tracing out the modes lying on the exterior. We consider the size 
of sphere to be much larger than the deSitter radius, $R\gg R_{\rm dS} = H^{-1}$,
where $H$ is the Hubble's constant. The entanglement entropy is 
computed using the usual methods and as expected the UV 
singularity comes from local physics (having no effect from 
non-locality) which we ignore. For very large spheres in 
four space-time dimensions, the finite piece has a term that 
goes as logarithm of the area of surface, we will focus on the coefficient of 
this piece. In odd-dimensions there are contribution to finite 
term which goes like area and a constant term. In odd-dimensions 
we therefore focus on the constant term. 

We compute the entanglement entropy for the non-local scalar field. 
To achieve this, we need to compute the density matrix 
by tracing out from modes lying outside the surface. When this 
spherical surface is taken all the way to the deSitter boundary then it 
is seen that the problem has a $SO(1,3)$ symmetry. The presence of 
this symmetry offers a helping hand in the computation of the 
density matrix and the entanglement entropy associated to it. 
The successful computation of the density matrix allows us to 
compute even Renyi entropies. 

We follow \cite{Callan:1994py} in defining entanglement entropy.
In a space-time divided in two regions by a closed surface 
$\Sg$ (the interior being $X$ and the exterior being $\bar{X}$),
then for a given field theory it is possible to write 
Hilbert space in to two parts. For example, if 
we have the Hilbert space of the field theory is denoted by $H$, 
then it is possible to do an approximate decomposition of $H = H_{X} \otimes H_{\bar{X}}$,
where $H_{X}$ contain modes localised inside the surface, while 
$H_{\bar{X}}$ contain modes outside surface. The basis in the two regions 
can then correspondingly be denoted by $\ket{\psi_n}_X$ and 
$\ket{\psi_n}_{\bar{X}}$. A generic state vector $\ket{\Psi}$ can then 
be expressed as a superposition 
\beq
\label{eq:psiSuperpose}
\ket{\Psi} = \sum_{n,m} C_{nm} \ket{\psi_n}_{X} \otimes \ket{\psi_m}_{\bar{X}} \, .
\eeq
The total density matrix then is given by
\beq
\label{eq:totDenmock}
\rho_{\rm tot} = \sum_{nm} \sum_{kl} 
C_{nm} C^*_{kl} \ket{\psi_n}{}_X\bra{\psi_k} 
\otimes \ket{\psi_m}{}_{\bar{X}}\bra{\psi_l} 
\eeq
On tracing out modes of either $X$ or $\bar{X}$ region 
one obtains the reduced density matrix 
\beq
\label{eq:redRhoX}
\rho_{X} = \sum_{nm} \sum_{kl} 
C_{nm} C^*_{kl} \ket{\psi_n}{}_X\bra{\psi_k} 
\sum_j {}_{\bar{X}}\bra{\psi_j}\ket{\psi_m}{}_{\bar{X}}\bra{\psi_l} \ket{\psi_j}_{\bar{X}} \, 
= \sum_{nk} \left(C C^\dagger\right)_{nk} \ket{\psi_n}{}_X \bra{\psi_k} \, ,
\eeq
where in performing simplification we have used ortho-normality of
states. This shows that the reduced density matrix in the basis 
$\ket{\psi_n}$ is given by $CC^\dagger$. This can be diagonalised 
to obtain its eigenvalues which we denote as $p_n$. Then the 
entropy of the system is given by,
\beq
\label{eq:vonNew}
S = - \sum_n p_n \log p_n = - {\rm Tr} \rho_{X} \log \rho_X \, .
\eeq
This is the definition of the entropy which we will be using 
in to compute the entanglement between modes lying 
in two regions.

In the case of deSitter space-time in flat slicing, we consider 
a given time-slice, where a closed surface $\Sg$ divides the region in to
interior and exterior: interior $X$ and exterior $\bar{X}$. 
This again results in a decomposition of the Hilbert 
space in to two parts. In four space-time dimensions 
in flat slicing of deSitter this surface $\Sg$ is a closed 
surface $S^2$ which divides the space-like hyper-surface in two 
regions. By tracing out modes either of exterior or interior one 
obtains the reduced density matrix whose eigenvalue are 
used in the computation of the entropy. 

In order to successfully implement this procedures for the 
case of deSitter we consider the deSitter space-time in flat slicing in four space-time 
dimensions:
\beq
\label{eq:ds4_flatslice}
{\rm d}s^2 = \frac{1}{H^2 \eta^2} \left[-{\rm d}\eta^2 + {\rm d}x_1^2 + {\rm d}x_2^2
+ {\rm d} x_3^3\right] \, ,
\eeq
where $H$ is the Hubble scale, $\eta$ is the conformal time and 
slicing is done at constant $\eta$. In the case when we have free massive (mass $m$) 
local scalar field theory, it is seen that the various contributions to the 
entanglement entropy can be structured in the following manner,
\bea
\label{eq:EEform}
S = c_1 \frac{{\cal A}}{\ep^2} + \log(\ep H) \left(
c_2 + c_3 A m^2 + c_4 A H^2
\right) + S_{\rm UV\, finite} \, ,
\eea
where $\ep$ is the UV-cutoff and ${\cal A}$ is the proper area. 
The first term is the well-known area
entropy \cite{Bombelli:1986rw,Srednicki:1993im}, which arises from the 
entanglement of particles residing near the surface. The terms proportional 
$c_2$ and $c_3$ are present in flat space-time, while the last term 
$c_4$ is dependent on curvature of bulk space. All the UV-divergent terms 
arise due to local effects, their coefficients will be same as in flat space-time. 
The UV-finite part contains effects from long-range correlation which has been the 
subject of investigation recently in various context 
\cite{Maldacena:2012xp,Kanno:2014lma,Iizuka:2014rua,Kanno:2014ifa,
Kanno:2014bma,Kanno:2016qcc,Kanno:2016gas,Vancea:2016tkt,
Kanno:2017dci,Choudhury:2017bou,Bhattacharya:2018yhm}.

In this paper we also study the UV-finite part of the entropy and its behaviour 
under the effect of non-locality. As noted in \cite{Maldacena:2012xp}
this UV-finite part contains information about the long-range 
correlations of the quantum state in deSitter space. Moreover, we 
expect that the long-distance part of the state to becomes time-independent. 
This is inferred by noticing that such long-range entanglement were established 
when these distances were of sub-horizon in size. 
Once they cross the horizon they tend to freeze and remains 
undisturbed by the evolution. As a result the long-distance 
piece of the entanglement is expected to be constant at late-times. 
This implies that if we fix a surface in co-moving coordinates, 
then proceeding to late times (which corresponds to limit $\eta\to0$)
it should be expected that the entanglement will be constant.
It has been noted in \cite{Maldacena:2012xp}
this is not true, in the sense that entanglement increases 
at long-distances. Moreover, as mass of field decreases the entanglement 
becomes further strong hinting that non-locality might have a crucial role to play. 

The finite part of EE then can be written as,
\beq
\label{eq:EEfiniteform}
S_{\rm UV-finite} = c_5 {\cal A} H^2 + \frac{c_6}{2} \log({\cal A}H^2)
+ {\rm finite} 
= c_5 \frac{A_c}{\eta^2} + c_6 \log \eta 
+ {\rm finite} \, ,
\eeq
where ${\cal A}$ is the proper area of the surface and $A_c$ is the 
area in co-moving coordinates (${\cal A} = A_c/(H\eta)^2$). The coefficient 
$c_6$ is contains information about the long-range 
entanglement of the state, which has been the subject of study in 
\cite{Maldacena:2012xp} and will also be under investigation 
in this paper. We will like to see the effect of non-locality on $c_6$. 

To achieve this aim we consider a spherical 
entangling surface $S^2$ ($x_1^2+x_2^2+x_3^2=R_c^2$) with radius $R_c\gg\eta$ 
implying that the surface is much larger than the horizon. 
Keeping $R_c$ fixed, the $\eta\to0$ limit leads to a surface on 
boundary which is left invariant by $SO(1,3)$ subgroup 
of $SO(1,4)$ dS isometry group. We expect that the coefficient 
$c_6$ to be respecting this symmetry. It is therefore better to adopt 
a co-ordinate system where this symmetry is more manifestly realised. 
This is two step process: first expressing deSitter in global co-ordinates 
where equal time slices are $S^3$, then secondly taking the entangling 
surface to be $S^2$ equator of $S^3$ \cite{Maldacena:2012xp}. 
Moreover, at $\eta=0$ any $S^2$ on the boundary can be mapped 
to equator of $S^3$ by exploiting the deSitter isometry. In the end, 
to order to regularise the problem the two-sphere is moved 
to a very late global time surface. 

\section {Mode functions}
\label{modeCom}

In this section we will compute the field modes for our non-local theory. 
These will be required in our computation of the entanglement entropy. 

The hyperbolic/open slicing of dS can be obtained by doing analytic continuation 
of sphere $S^4$ sliced by $S^3$ \cite{Bucher:1994gb,Sasaki:1994yt}. The metric in 
Euclidean signature is given by, 
\beq
\label{eq:eucmetS4}
{\rm d}s_E^2 = H^{-2} \left(
{\rm d}\tau_E^2 + \cos^2 \tau_E \left(
{\rm d}\rho_E^2 + \sin^2 \rho_E^2 {\rm d}\OM_2^2
\right)
\right)
\eeq
Analytic continuation of this leads to Lorentzian manifold which can be 
divided into three parts. These Lorentizian manifolds are 
related to Euclidean in the following way:
\begin{align}
\label{eq:LorCharts}
&
R:
\begin{cases}
\tau_E = \pi/2 - i t_R & t_R\geq0 \, ,\\
\rho_E = -i r_R        &   r_R\geq0 \, , \\
\end{cases} \notag \\
&
C: 
\begin{cases}
\tau_E = t_C & -\pi/2\leq t_C\leq\pi/2 \, ,\\
\rho_E = \pi/2 - i r_C        &   -\infty<r_C<\infty\, , \\
\end{cases} \notag \\
&
L: 
\begin{cases}
\tau_E = -\pi/2 + i t_L & t_L\geq0 \, ,\\
\rho_E = -i r_L        &   r_L\geq0 \, , \\
\end{cases} 
\end{align}
The metric in each region is given by,
\bea
\label{eq}
{\rm d}s_R^2 = H^{-2} \left(
-{\rm d}t_R^2 + \sinh^2 t_R \left(
{\rm d}r_R^2 + \sinh^2 r_R^2 {\rm d}\OM_2^2
\right)
\right) \, ,
\notag \\
{\rm d}s_C^2 = H^{-2} \left(
{\rm d}t_C^2 + \cos^2 t_C \left(
-{\rm d}r_C^2 + \cosh^2 r_C^2 {\rm d}\OM_2^2
\right)
\right) \, ,
\notag \\
{\rm d}s_L^2 = H^{-2} \left(
-{\rm d}t_L^2 + \sinh^2 t_L \left(
{\rm d}r_L^2 + \sinh^2 r_L^2 {\rm d}\OM_2^2
\right)
\right) \, ,
\eea
In the case of massive scalar as studied in \cite{Maldacena:2012xp}
(or spinors \cite{Kanno:2016qcc}) one can solve the equation of motion 
for the mode function in the $L$ and $R$ region of dS. In the case when 
non-locality is present, one gets a modification for the scalar field action 
as given in eq. (\ref{eq:NLact}). This leads to following equation of motion 
for the scalar field
\beq
\label{eq:EQMscalar}
\left(-\Box + m^2 + \frac{\lam^2}{-\Box} \right) \phi(t,r,\OM)=0 \, ,
\eeq
where $\Box = \nb_\mu \nb^\mu$. The field $\phi$ can be decomposed 
into various modes lying on $R$ and $L$ part of the dS. 
The mode-functions also satisfy the eq. (\ref{eq:EQMscalar}). 
The operator in the eq. (\ref{eq:EQMscalar}) being 
non-local in nature causes some complications. However, 
an interesting observation leads to simplification 
\cite{Kajuri:2018wow,Narain:2018rif,Narain:2018nfh}, as
the operator can be factored. We will exploit this feature 
again to obtain mode-functions for the case presented here. 
If we have mode-function $u_{-}(t,r,\OM)$ and $u_{+}(t,r,\OM)$
satisfying
\beq
\label{eq:phi1phi2}
(-\Box + r_{-}^2) u_{-}(t,r,\OM)=0 \, ,
\hspace{5mm}
(-\Box + r_{+}^2) u_{+}(t,r,\OM)=0 \, .
\eeq
where 
\beq
\label{eq:rpm}
r_{-}^2 = (m^2 - \sqrt{m^4 - 4\lam^2})/2\, ,
\hspace{5mm} 
r_{+}^2= (m^2 + \sqrt{m^4 - 4\lam^2})/2 \, ,
\eeq
then we have that 
\beq
\label{eq:UmodeFun}
u(t,r,\OM) = A_{-} u_{-}(t,r,\OM) + A_{+} u_{+}(t,r,\OM) \, 
\eeq
satisfies the eq. (\ref{eq:EQMscalar}) while 
\beq
\label{eq:ABcoeff}
A_{-} = - r_{-}^2/(r_{+}^2 - r_{-}^2) \, ,
\hspace{5mm}
A_{+} = r_{+}^2/(r_{+}^2 - r_{-}^2) \, .
\eeq
Moreover, in the limit $\lam\to0$ (locality limit) the mode-function $u(t,r,\OM)$ 
reduces to mode-function for local massive scalar field as $A\to0$. We will 
make use of this knowledge to work out the mode-functions for our non-local 
case. This also implies that scalar field $\phi$ can be expressed 
as a linear combination $\phi_{-}$ and $\phi_{+}$: $\phi = \sum_s A_s \phi_s$
where $s=-,+$. 

This simplifies the problem very much where now one has to 
solves the mode function for each part. The equation of motion 
for the mode function in four dimensions for each part is given by:
\beq
\label{eq:modefunc}
\biggl[
\frac{1}{\sinh^3 t} \frac{\pt}{\pt t} \sinh^3 t \frac{\pt}{\pt t}
- \frac{1}{\sinh^2 t} \mathbb{L}_{H^3}^2 + \frac{9}{4} - \nu_{s}^2
\biggr] u_{s} (t,r,\OM) =0 \, ,
\eeq
where $\mathbb{L}_{H^3}^2$ is the Laplacian on the unit hyperboloid and 
\beq
\label{eq:nupm}
\nu_{s}^2 = \frac{9}{4} - \frac{r_{s}^2}{H^2} \, .
\eeq
The wave-functions are labeled by the quantum numbers corresponding to the 
Casimir of $H^3$ and angular momentum on $S^2$. 
\beq
\label{eq:wavemodes}
u^{s}_{plm} \sim \frac{H}{\sinh t} \chi_{s,p}(t) Y_{plm}(r,\OM) \, ,
\hspace{5mm}
- \mathbb{L}_{H^3} Y_{plm} = (1+p^2) Y_{plm} \, .
\eeq
where $Y_{plm}$ are the eigenfunction on the hyperboloid (analogous to 
spherical harmonics) \cite{Sasaki:1994yt}. The time-dependent part of the 
wave-function is contained in $\chi_{s,p}(t)$ (modulo the overall factor of 
$1/\sinh t$). The positive frequency wave-function corresponding to 
Euclidean vacuum is chosen by demanding them to be analytic when 
continued to lower hemisphere. These wave-function of the $\pm$-parts 
will have support on both the $R$ and $L$ regions of dS. They are 
given by,
\begin{align}
\label{eq:modechi}
\chi_{p,\sg}^s = 
\begin{cases}
&
\frac{1}{2\sinh(\pi p)} \biggl[
\frac{e^{\pi p} - i \sg e^{-i \pi \nu_{s}}}{\G(\nu_s + i p +1/2)} 
P^{ip}_{\nu_s -1/2}(\cosh t_R) 
- \frac{e^{-\pi p} - i \sg e^{-i \pi \nu_{s}}}{\G(\nu_s - i p +1/2)} 
P^{-ip}_{\nu_s -1/2}(\cosh t_R) 
\biggr] \, ,\\
&
\frac{\sg}{2\sinh(\pi p)} \biggl[
\frac{e^{\pi p} - i \sg e^{-i \pi \nu_{s}}}{\G(\nu_s + i p +1/2)} 
P^{ip}_{\nu_s -1/2}(\cosh t_L) 
- \frac{e^{-\pi p} - i \sg e^{-i \pi \nu_{s}}}{\G(\nu_s - i p +1/2)} 
P^{-ip}_{\nu_s -1/2}(\cosh t_L) 
\biggr] \, .
\end{cases}
\end{align}
The parameter $\sg=\pm1$ and $s=+,-$. For each $\sg$ and $s$ the top line gives the 
mode-function on the $R$-hyperboloid while the lower line is the 
mode-function on $L$-hyperboloid. These two solutions for each values 
of $\sg$ and $s$ form a basis on the $R$ and $L$ region of the dS in terms of 
which the scalar field $\phi$ can be expanded. The operator $\phi$ field  
will consist of two parts:
\beq
\label{eq:phiexp}
\hat{\phi}(x) = \int {\rm d}p \sum_{s,\sg,l,m} \left[
A_s a_{s,\sg,plm} u_{s,\sg,plm} + A_s a^{\dagger}_{s,\sg,plm} \bar{u}_{s,\sg,plm}
\right]  \, ,
\eeq
where the functions $u$'s contain $x$ dependence. In the next section we will use 
this decomposition to compute the density matrix. 

\section{Density matrix}
\label{denmat}

To compute density matrix 
one has to trace out degree of freedom in either $R$ or $L$ region of dS, to do 
this it is better to do a change of basis which has support on either $R$ or $L$ 
region. For $R$ region we choose the basis function to be $P^{ip}_{\nu^{s}-1/2}(\cosh t_R)$
and $P^{-ip}_{\nu^{s}-1/2}(\cosh t_R)$, and zero in the $L$ region. These 
are positive and negative frequency wave functions in the $R$ region 
for both values of $s=+,-$. For the $L$ region same thing holds, the basis 
is given by $P^{ip}_{\nu^{s}-1/2}(\cosh t_L)$
and $P^{-ip}_{\nu^{s}-1/2}(\cosh t_L)$ and zero in $R$ region. 
These should be properly normalised using the Klein-Gordon norm. 
The original mode-function in eq. (\ref{eq:modechi}) can then be written 
in terms of new basis by exploiting matricial form 
\begin{align}
\label{eq:basisCH}
\begin{cases}
&
\chi^\sg_{s} = N_P^{-1} \sum_{q=R,L} \left[
(\al_s)^\sg_q P^q_{s} + (\bt_s)^\sg_q \bar{P}_s^q
\right] \, \\
&
\bar{\chi}^\sg_s = N_P^{-1} \sum_{q=R,L} \left[
(\bar{\bt}_s)^\sg_q P^q_{s} + (\bar{\al}_s)^\sg_q \bar{P}_s^q
\right] \,\\
\end{cases}
\Rightarrow
\chi^I_s = (M_s)^I{}_J P^J_s \, N_p^{-1} \, ,
\end{align}
where $N_p^{-1}$ is the normalisation factor including the factor 
of $1/\sinh(\pi p)$, $\sg= \pm1$, 
$P^{R,L}_s \equiv P^{ip}_{\nu^{s}-1/2}(\cosh t_{R,L})$
and $\chi^I_s = \{\chi^\sg_s, \bar{\chi}^\sg_s\}$. In should be 
specified that the repeated $s$ index here doesn't imply summation 
over $s$. The matrix $\al_s$ and $\bt_s$ follows from 
eq. (\ref{eq:modechi}) and are given by, 
\begin{align}
\label{eq:albtRL}
&
(\al_s)^\sg_R = \frac{e^{\pi p} - i \sg e^{-i \pi \nu_s}}{\G(\nu_s + i p +1/2)} \, ,
& 
(\bt_s)^\sg_R = - \frac{e^{-\pi p} - i \sg e^{-i \pi \nu_s}}{\G(\nu_s - i p +1/2)} \, ,
\notag
\\
&
(\al_s)^\sg_L = \frac{\sg(e^{\pi p} - i \sg e^{-i \pi \nu_s})}{\G(\nu_s + i p +1/2)} \, ,
&
(\bt_s)^\sg_L = - \frac{\sg(e^{-\pi p} - i \sg e^{-i \pi \nu_s})}{\G(\nu_s - i p +1/2)} \, .
\end{align}
While the matrix $M_s$ and $P^J_s$ follows from eq. (\ref{eq:basisCH}) 
and are given by
\beq
\label{eq:Mmat}
(M_s)^I{}_J = \left(
\ba{c c}
(\al_s)^\sg_q & (\bt_s)^\sg_q \vspace{3mm}\\
(\bar{\bt}_s)^\sg_q & (\bar{\al}_s)^\sg_q 
\ea
\right) \, ,
\hspace{5mm}
P^J_s = \{P^R_s,P^L_s,\bar{P}^R_s, \bar{P}^L_s\} \, 
\eeq
respectively. In this new notation the scalar field $\phi$ undergoes a basis change.
In short-hand notation it can be written as,
\beq
\label{eq:phinew}
\phi(t) = \sum_{s=+,-} A_s (a_s)_I \chi^I_s 
= N_p^{-1} \sum_{s=+,-} A_s (a_s)_I (M_s)^I{}_J \, P^J_s \, ,
\hspace{5mm}
(a_s)_I = \{(a_s)_\sg, (a_s)^\dagger_\sg \} \, ,
\eeq
where the vacuum is defined by $(a_s)_\sg \ket{\Psi_s}=0$.
In order to trace-out modes in the $R$ or $L$ region, one has to do 
Bogoliubov transformation to change the basis. In the new basis the 
creation and annihilation operators will be different $(b_s)_J
= \{(b_s)_R, (b_s)_L,(b_s)_R^\dagger,(b_s)^\dagger_L\}$, while 
$\phi(t) = \sum_{s} \, (b_s)_J P^J_s$. We can define a matrix
$(\mathbb{M}_s)^{}_J = A_s (M_s)^I{}_J$, where summation 
over $s$ is not implied. By using the matrix transformation one 
can express the operators $a$'s in terms of $b$'s. This is given by,
\beq
\label{eq:asig}
(a_s)_J = (b_s)_I \left(\mathbb{M}_s^{-1}\right)^I{}_J \, .
\eeq
As the entries of matrix $M_s$ given in eq. (\ref{eq:Mmat}) 
which are itself $2\times2$ matrix, 
so one can make use of inversion formula to obtain the 
inversion of $\mathbb{M}_s$. The entries of $\mathbb{M}_s^{-1}$ 
will be given by,
\beq
\label{eq:Minv}
\mathbb{M}_s^{-1} = \left(
\ba{c c}
(\g_s)^\sg_q & (\de_s)^\sg_q \vspace{3mm} \\
(\bar{\de}_s)^\sg_q & (\bar{\g}_s)^\sg_q 
\ea
\right) \, ,
\eeq
where the entries $\{\g,\de,\bar{\de},\bar{\g}\}$ can be expressed 
in terms of entries of $\mathbb{M}$ as follows
\beq
\label{eq:Minvent}
\g_s = A_s^{-1} (\al_s - \bt_s \bar{\al_s}^{-1} \bar{\bt}_s)^{-1} \, , 
\hspace{5mm}
\de_s = - A_s^{-1} \al_s^{-1} \bt_s \bar{\g}_s \, .
\eeq
This immediately gives the expression for the operators 
$a_{s}$ appearing in eq. (\ref{eq:phiexp}) in terms of 
new operators $b_s$. 
\beq
\label{eq:asg}
(a_s)_\sg = \sum_{q=R,L} \left[
(b_s)_q (\g_s)_\sg^q 
+ (b_s)^\dagger (\bar{\de}_s)_\sg^q 
\right] \, ,
\eeq
where again there is no summation over $s$. 
In this sense one can see the Bunch-Davis vacuum as a Bogoliubov 
transformed vacua of $R$ and $L$ region. 
\beq
\label{eq:BDtrans}
\ket{\Psi_s} = \exp\left[
\frac{1}{2} \sum_{i,j=R,L} (m_s)_{ij} (b_s)_i^\dagger (b_s)_j^\dagger 
\right] \ket{R_s} \ket{L_s} \, ,
\eeq
The matrix $(m_s)_{ij}$ can be determined by noting 
$(a_s)_\sg\ket{\Psi_s}=0$, which translates in to the 
condition $(m_s)_{ij} (\g_s)_{j\sg} + (\bar{\de}_s)_{i\sg}=0$. On inversion 
this gives,
\beq
\label{eq:mij}
(m_s) = - (\bar{\de}_s)(\g_s)^{-1}
= - \frac{\G(\nu_s - ip+1/2)}{\G(\nu_s + ip+1/2)}
\frac{2e^{i\pi \nu_s}}{e^{2\pi p} + e^{2i\pi\nu_s}}
\left(
\ba{c c}
\cos \pi\nu_s & i \sinh p \pi \\
i \sinh p\pi & \cos \pi\nu_s
\ea
\right) \, .
\eeq
It should be mentioned here that the matrix $m_s$ doesn't have 
dependence on $A_s$ as it cancels out. 
This matrix has a phase factor which is unimportant and can be
absorbed in the definition of operators in new basis. Expressing the 
phase factor as $\exp(i\ta_s)$ (phase corresponding for each part of $\phi$), 
it is seen that one has \cite{Maldacena:2012xp}
\beq
\label{eq:mijphase}
(m_s) = e^{i\ta_s} 
\frac{\sqrt{2} e^{-p\pi}}{\sqrt{\cosh 2\pi p + \cos 2\pi\nu_s}}
\left(
\ba{c c}
\cos \pi\nu_s & i \sinh p \pi \\
i \sinh p\pi & \cos \pi\nu_s
\ea
\right) \, .
\eeq
In this form the degree of freedom in the $R$ and $L$ region are 
still mixed with each other as a result it is difficult to trace out 
modes of either $R$ or $L$. This demands a further transformation 
and an introduction of new set of operators 
$(c_s)_R^\dagger$, $(c_s)_L^\dagger$, $(c_s)_R$ and $(c_s)_L$. 
Then the original vacuum $\ket{\Psi_s}$ can be written as
\beq
\label{eq:psinew}
\ket{\Psi} = \ket{\Psi_+} \bigotimes \ket{\Psi_{-}}
= \left(
\sum_{n^{+}_R,m^{+}_L} d_{n^{+}_R,m^{+}_L} \ket{n^{+}_R} \ket{m^{+}_L} 
\right) \bigotimes 
\left(
\sum_{n_R^{-},m_L^{-}} d_{n^{-}_R,m^{-}_L} \ket{n^{-}_R} \ket{m^{-}_L} 
\right)
\, ,
\eeq
where $d_{n^{+}_R,m^{+}_L}$, $d_{n^{-}_R,m^{-}_L}$ are 
coefficients while the new set of operators 
$(c_s)_R^\dagger$, $(c_s)_L^\dagger$ satisfy 
\beq
\label{eq:cRcL}
(c_s)_R^\dagger \ket{n^s_R} = \sqrt{n^s_R+1} \ket{n^s_R+1} \, ,
\hspace{5mm}
(c_s)_L^\dagger \ket{n^s_L} = \sqrt{n^s_L+1} \ket{n^s_L+1} \, 
\eeq
respectively. The new vacua is defined using the new operators 
$(c_s)_R$ and $(c_s)_L$ in the following manner
\beq
\label{eq:newVac}
(c_s)_R \ket{R_s^\prime} =0 \, ,
\hspace{5mm}
(c_s)_L \ket{L_s^\prime} =0 \, .
\eeq
The new set of operators can be expressed in terms of 
$b$ and $b^\dagger$ by making use of linear transformation
\beq
\label{eq:cinb}
(c_s)_R = u_s (b_s)_R + v_s (b_s)_R^\dagger \, ,
\hspace{5mm}
(c_s)_L = \bar{u}_s (b_s)_L + \bar{v}_s (b_s)_L^\dagger \, ,
\eeq
under the constraint $\lvert u_s \rvert^2 - \lvert v_s \rvert^2=1$. If we further impose the condition
that $(c_R^s) \ket{\Psi_s} = \g_s (c_L^s)^\dagger \ket{\Psi_s}$
and $(c_L^s) \ket{\Psi_s} = \g_s (c_R^s)^\dagger \ket{\Psi_s}$, then 
using eq. (\ref{eq:cinb}) and (\ref{eq:BDtrans}) one can solve for $u_s$, $v_s$ and 
$\g_s$, where $\g_s$ is given by \cite{Maldacena:2012xp,Kanno:2014lma}
\beq
\label{eq:gampm}
\g_s = i \frac{\sqrt{2}}
{
\sqrt{\cosh 2\pi p + \cos 2\pi\nu_s}
+ \sqrt{\cosh 2\pi p + \cos 2\pi\nu_s+2}
} \, .
\eeq
These set of constraints and conditions also requires 
that the coefficients $d_{n^{+}_R,m^{+}_L}$, $d_{n^{-}_R,m^{-}_L}$ 
appearing in eq. (\ref{eq:psinew}) to satisfy the following 
conditions
\begin{align}
\label{eq:dcond}
& \sqrt{n^s_R+1} \, d_{n^s_R+1,m^s_L}
= \g_s \sqrt{m_L^s} d_{n^s_R,m^s_L-1} \, , 
\notag \\
& \sqrt{m^s_L+1} d_{n^s_R,m^s_L+1}
= \g_s \sqrt{n_R^s} d_{n^s_R-1,m^s_L} \, .
\end{align}
The only possibility for these two conditions to be 
simultaneously satisfied is when
\beq
\label{eq:dnm}
d_{n^s_{R},m^s_{L}} = \lvert\g_s\rvert^{n_R^s} 
\sqrt{1-\lvert\g_s\rvert^2} \de_{n^s_R,m^s_L} \, .
\eeq
This particular for the coefficients $d_{n^{+}_R,m^{+}_L}$, $d_{n^{-}_R,m^{-}_L}$
when plugged back in eq. (\ref{eq:psinew}) and making use of 
eq. (\ref{eq:cRcL}) results in the following form 
of the Bunch-Davis vacuum
\begin{align}
\label{eq:BDvacC}
\ket{\Psi_s}
& = \sqrt{1- \lvert\g_s\rvert^2} \sum_{n^s}
\lvert\g_s\rvert^{n^s} \ket{n^s}_R\ket{n^s}_L
\notag \\
& = \sqrt{1- \lvert\g_s\rvert^2} \exp[\lvert\g_s\rvert (c_R^s)^\dagger (c_L^s)^\dagger] 
\ket{R_s^\prime}\ket{L_s^\prime} \, .
\end{align}
Once we have obtained this then now 
it is now easy to integrate out modes lying either on $R$ or $L$ region.
The total density matrix is given by,
\beq
\label{eq:dentot}
\rho_{\rm tot} = \ket{\Psi}\bra{\Psi}
= \ket{\Psi_+}\bra{\Psi_+} \otimes \ket{\Psi_{-}}\bra{\Psi_{-}}
= \rho^{+}_{\rm tot} \otimes \rho^{-}_{\rm tot}
\eeq
where 
\beq
\label{eq:rhoStot}
\rho^s_{\rm tot} = (1- \lvert\g_s\rvert^2)
\sum_{n_s,m_s} \lvert\g_s\rvert^{n^s} \lvert\g_s\rvert^{m^s} 
\ket{n^s} {}_R \bra{m^s} \ket{n^s} {}_L \bra{m^s} \, .
\eeq
Integrating out the $R$ part leaves us with the reduced density matrix $\rho_{\rm red}$.
The reduced density can be computed for both $+$ and $-$, which is 
correspondingly given by $\rho^{+}_{\rm red}$ and $\rho^{-}_{\rm red}$
respectively. 
\begin{align}
\label{eq:redDen}
\rho_{\rm red}
&= \rho^{+}_{\rm red} \otimes \rho^{-}_{\rm red}
= (1 - \lvert\g_{-}\rvert^2)(1 - \lvert\g_{+}\rvert^2) \left(\sum_n \lvert\g_{-}\rvert^{2n} \ket{n}{}_L\bra{n} \right) 
\bigotimes 
\left(
\sum_m \lvert\g_{+}\rvert^{2m} \ket{m}{}_L\bra{m} 
\right)
\notag \\
&= 
(1 - \lvert\g_{-}\rvert^2)(1 - \lvert\g_{+}\rvert^2) \sum_n \lvert\g_{-} \g_{+}\rvert^{2n} \ket{n}{}_L\bra{n} \, .
\end{align}
From this we notice that the density matrix $\rho_{\rm red}$ is diagonal. 
This reflects from the fact that there is no entanglement among the 
states with different $SO(1,3)$ quantum numbers. It should be mentioned here 
that as the non-local system can be expressed as a local system of two 
decoupled scalars so the above computations can be repeated for the decoupled system. 
In this case again we will find that the total density matrix will be a product of 
density matrix corresponding to the two decoupled scalars. This is expected as the 
lagrangian of the two theories are related to each other. 

Once the density matrix is obtained it is easy to notice the eigenvalues of the density matrix. 
They are given by,
\beq
\label{eq:eigenRho}
p_n = (1 - \lvert\g_{-}\rvert^2)(1 - \lvert\g_{+}\rvert^2) \lvert\g_{-} \g_{+}\rvert^{2n} 
= p_n^{-} p_n^{+}\, .
\eeq
From this one can immediately compute the entanglement entropy 
using $S = - \sum_n p_n \log p_n$. This gives,
\bea
\label{eq:EE}
&&
S(p,\nu_{+},\nu_{-}) = - (1 - \lvert\g_{-}\rvert^2)(1 - \lvert\g_{+}\rvert^2)
(1- \lvert\g_{-}\rvert^2 \lvert\g_{+}\rvert^2)^{-1}
\notag \\
&&
\times
\biggl[
\log\{(1 - \lvert\g_{-}\rvert^2)(1 - \lvert\g_{+}\rvert^2) \}
+ \frac{\lvert\g_{-}\rvert^2 \lvert\g_{+}\rvert^2}{(1- \lvert\g_{-}\rvert^2 \lvert\g_{+}\rvert^2)}
\log(\lvert\g_{-}\rvert^2 \lvert\g_{+}\rvert^2)
\biggr] \, .
\eea
This has symmetry $\g_{-} \leftrightarrow \g_{+}$. 
The final entropy is computed by summing eq. (\ref{eq:EE}) over all the states. 
The quantity $c_6$ which contains information of long range correlations 
as mentioned in eq .(\ref{eq:EEfiniteform}) can be obtained by integrating 
over $p$ and the volume integral over the hyperboloid. In other words, we 
use density of states on the hyperboloid. The full entropy is
\beq
\label{eq:Snufull}
S(\nu_{+}, \nu_{-}) = \mathbb{V}_{H^3} \int {\rm d}p {\cal D}(p) 
S(p,\nu_{+},\nu_{-}) \, ,
\eeq
where ${\cal D}(p) = p^3/(2\pi^2)$ is the density of states for the hyperboloid 
\cite{Bytsenko:1994bc} and $\mathbb{V}_{H^3}$ is the volume of hyperboloid. 
This volume is infinite as the entangling surface is taken all the way up 
$\eta=0$. It is regularised by applying a radial cutoff which implies putting 
the entangling surface at finite time. As we are interested in the coefficient of
$\log \eta$ in the EE, so the precise way of implementing the cutoff at 
large volumes doesn't matter. The final answer for the coefficient $c_6$ 
of the $\log \eta$ in EE is given by \cite{Maldacena:2012xp},
\beq
\label{eq:EElogterm}
S_{\rm intr} \equiv c_6 = \frac{1}{\pi} \int_0^\infty
{\rm d}p \, p^2 S(p, \nu_{+}, \nu_{-}) \, ,
\eeq
where $S(p,\nu_{+}, \nu_{-})$ is given in eq. (\ref{eq:EE}). 
At this point we notice that there system has two parameters: $m$
and $\lam$ (apart from $H$) which create an interesting interplay.

\subsection{Locality limit $\lam\to0$}
\label{lam0lim}

In the case when there is no non-locality ($\lam\to0$) we notice 
that $r_{-}^2\to0$ and $r_{+}^2\to m^2$. In this limit then 
we have $\nu_{-} \to 3/2$ while $\nu_{+}\to \sqrt{9/4 - m^2/H^2}$. 
In this limit $A_{-}\to0$, as a result the mode-function for non-local case will 
reduce to massive local case. However, it is noticed 
that the contribution of this $(-)$ piece doesn't disappear 
from the entropy, as $\lvert\g_{-}\rvert$ doesn't vanish. 
In this case although $r_{-}^2$ goes to zero, still the corresponding 
density matrix is not unity. 
In this limit we have $\lvert\g_{-}\rvert$ given by,
\begin{align}
\label{eq:gamMlam0}
\lvert\g_{-}\rvert &= \frac{\sqrt{2}}
{\sqrt{\cosh 2\pi p -1} + \sqrt{\cosh 2\pi p +1}} = e^{-p\pi} \, ,
\notag \\
\lvert\g_{+}\rvert & = \frac{\sqrt{2}}
{\sqrt{\cosh 2\pi p + \cos 2\pi\nu_{+}}
+ \sqrt{\cosh 2\pi p + \cos 2\pi\nu_{+}+2} } \, .
\end{align}
It should be noticed that even though we are considering the 
locality limit ($\lam\to0$), still $\g_{-}\neq1$ (happens only for $p=0$). 
This implies that the eigenvalues of the density 
matrix $\rho^{-}_{\rm red}$ are $p^{-}_n = 
(1 - e^{-2p\pi}) e^{-2np\pi}$. These eigenvalues are equal to 
unity only for some special values of $p$. This is 
interesting in the sense that even though the non-locality 
is slowly removed it still leaves an imprint behind on the density matrix. 
This can also be understood from the fact the local system 
of two decoupled scalars stated in eq. (\ref{eq:TransLagFT}) 
reduces in the limit $\lam\to 0$ to a decoupled system of 
massless and a massive scalar field. The massless scalar 
give rise to an additional contribution to the entanglement. 
In this sense our non-local system is different from the 
local system of a single scalar field considered in 
\cite{Maldacena:2012xp}, as in our case there is a presence 
of an additional massless scalar field.

\begin{figure}[h]
\centerline{
\vspace{0pt}
\centering
\includegraphics[width=4.5in,height=3in]{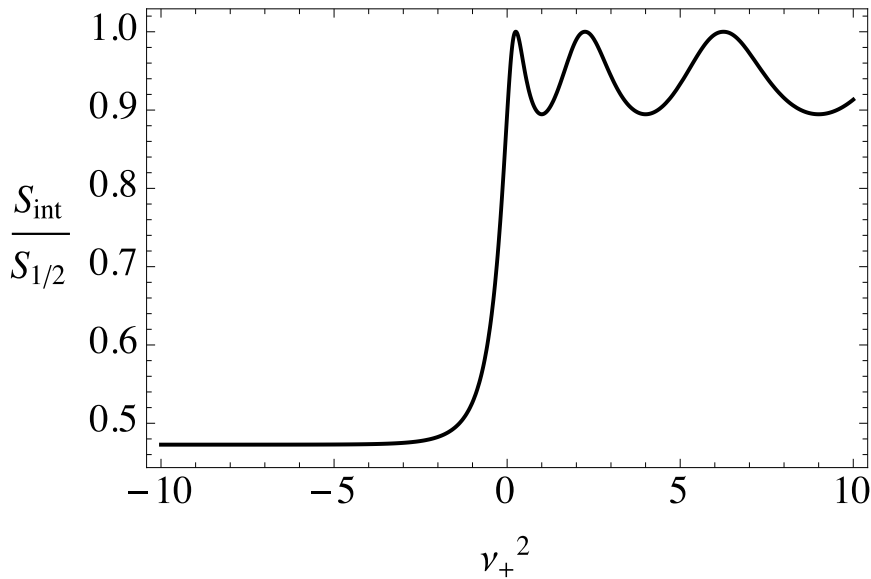}
}
\vspace{-3mm}
\caption[]{
Plot of entropy $S_{\rm int}/S_{{\rm int},1/4}$ for the case $\lam\to0$. 
In this case $r_{+}^2\to m^2$ and $r_{-}^2\to0$. 
We plot this against $\nu_{+}^2 = 9/4 - m^2/H^2$ (where we take $H=1$).
We have plotted the entropy for both physical and tachyonic mass
indicating the behaviour of entropy for physical and tachyonic mass. 
}
\label{fig:dsEElam0}
\end{figure}

The entropy $S(p,\nu_{+},\nu_{-})$ computed using 
the $\g_{-}$ and $\g_{+}$ given in eq. (\ref{eq:gamMlam0})
differs from the case of massive local scalar field
and matches only for the case of $p=0$ (when $\g_{-} \to 1$). 
We consider it as an imprint of non-locality. 
In figure \ref{fig:dsEElam0} we plot the entropy for the case of 
local field theory ($\lam\to0$). In this case we also notice some 
oscillations in the behaviour of entropy. This is normal 
as it is also true for the case of massive local scalar field 
theory if $\nu^2>9/4$. This is the tachyonic regime \cite{Maldacena:2012xp}. 
%
%

\subsection{Massless limit $m\to0$}
\label{m0}

In this section we consider the massless case ($m\to0$ limit). In this 
case $r_{-}^2 \to -i \lam$ while $r_{+}^2 \to i\lam$
(or in the case with positive sign of $\lam^2$ 
one will have $r_{-}^2 \to -\lam$ and $r_{+}^2 \to \lam$).
This gives the corresponding $\nu_{\pm}$ to be 
\beq
\label{eq:nusm0}
\nu_{+}^2 = \frac{9}{4} + \frac{\sqrt{-\lam^2}}{H^2} \, ,
\hspace{5mm}
\nu_{-}^2 = \frac{9}{4} - \frac{\sqrt{-\lam^2}}{H^2} \, .
\eeq
This then gives the corresponding $\g_\pm$ following 
eq. (\ref{eq:gampm}). In the massless case $\nu_\pm$ can be written in 
complex form (if $\lam^2 \to -\lam^2$ then $\nu_\pm^2$ will 
be real). The expression for $\g_\pm$ involves $\cos 2\pi \nu_\pm$. 
This is complex for complex $\nu_\pm$.  Although $\g_\pm$ is 
complex but the quantity that enters the definition of the entropy is 
the absolute value of $\g_\pm$, which is real. The expressions 
however are very complicated. It is worthwhile to investigate how the 
entropy behaves as a function of non-locality strength $\lam$. 
We consider two cases of theory given in eq. (\ref{eq:NLact})
which are related to each other by transformation $\lam^2\to -\lam^2$.
In the later we have a tachyon while in former we have complex 
poles. In the case when we have complex mass poles it is 
seen that the entropy first increases 
then decays exponentially. In the case when we have 
tachyon in theory then the entropy is seen to exhibits a oscillatory behaviour
which is expected even for massive local theories for 
tachyon field. These two situations can be better understood by 
considering the corresponding local theory stated in 
eq. (\ref{eq:TransLagFT}). In the massless case it is noticed from 
eq. (\ref{eq:kpConst}) that $\kp$ has two simple solutions $\pm1$. 
This decoupled system of two scalars will then give rise to corresponding
entanglement entropy agreeing with above case. So the case of 
decaying entanglement will correspond to scalar with complex 
masses in the decoupled system while the oscillating case will have a 
scalar with tachyonic mass. 
%
\begin{figure}[h]
\centerline{
\vspace{0pt}
\centering
\includegraphics[width=4.5in,height=3in]{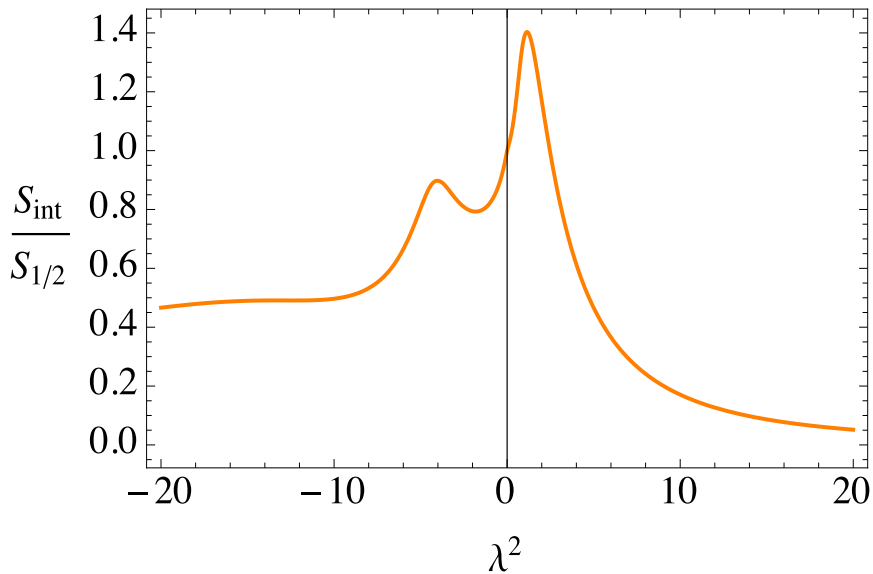}
}
\vspace{-3mm}
\caption[]{
Plot of entropy $S_{\rm int}/S_{{\rm int},1/4}$ for the case $m\to0$. 
We plot this against $\lam^2$ (where we take $H=1$).
This takes care of both $\pm\lam^2$.
In either case entanglement entropy is seen to decrease as 
non-locality strength increases. 
}
\label{fig:dsEEm0}
\end{figure}
%

\subsection{$m\neq0$ and $\lam\neq0$}
\label{m0lam0}

\begin{figure}[h]
\centerline{
\vspace{0pt}
\centering
\includegraphics[width=3.2in,height=2.2in]{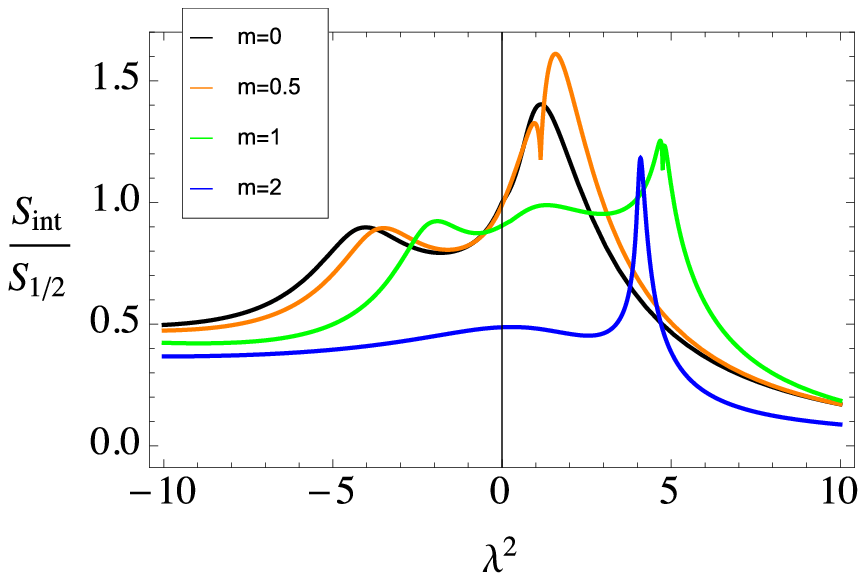}
\hspace{2mm}
\includegraphics[width=3.2in,height=2.1in]{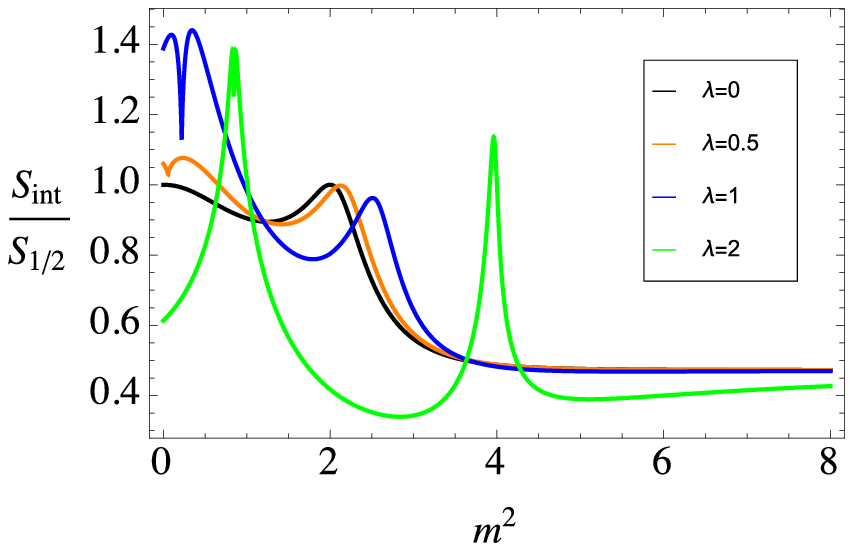}
}
\vspace{-3mm}
\caption[]{
Plot of entropy $S_{\rm int}/S_{{\rm int},1/4}$ for varying 
$\lam$ and $m$ (where we take $H=1$). It is seen that 
for small mass the entropy first increases then decreases 
as non-locality $\lam$ increases.
}
\label{fig:dsEEmlam}
\end{figure}

In the case when both $m$ and $\lam$ are non-zero then the 
entropy gets a more richer structure as there are two parameters. 
In the paper by \cite{Maldacena:2012xp} it was noticed that 
in local field theory as the mass decreases the long-range 
entanglement increases, which is a feature of deSitter. 
We decided to plot these cases to see the behaviour of 
entanglement entropy with varying $\lam$ and $m$. 
This behaviour is depicted in figure \ref{fig:dsEEmlam}. 
%
%

\section{R\'enyi entropy and negativity}
\label{RenNeg}

\begin{figure}[h]
\centerline{
\vspace{0pt}
\centering
\includegraphics[width=4.5in,height=3in]{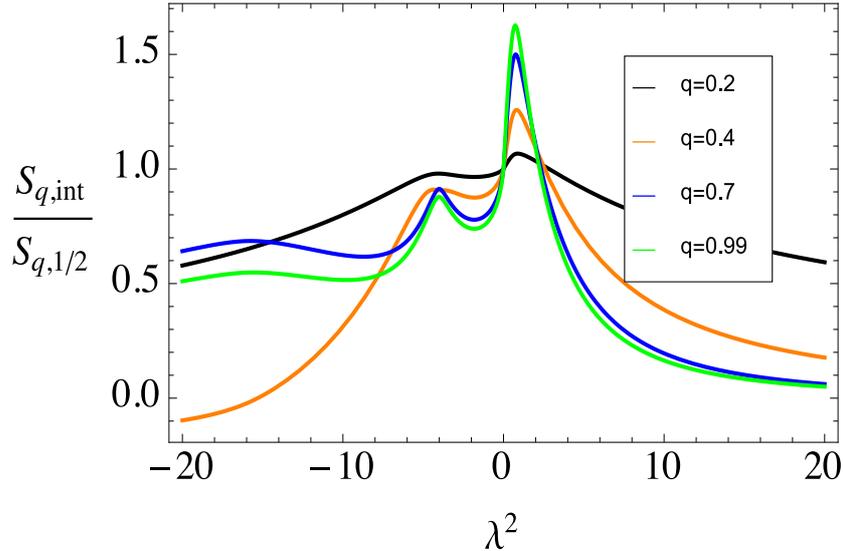}
}
\vspace{-3mm}
\caption[]{
Plot of R\'enyi entropy $S_{{\rm q, int}}/S_{{\rm q, int},1/4}$ for varying 
$\lam^2$ (where we take $H=1$). We plot the entropy for the 
case of $q=0.2$, $q=0.4$, $q=0.7$ and $q=0.99$. We consider the 
case of massless non-local theory. 
}
\label{fig:dsEERenM0}
\end{figure}

From the density matrix and eigenvalues one can also define
the corresponding R\'enyi-entropy which is defined as
\beq
\label{eq:rendef}
S_q = \frac{1}{1-q} \log {\rm Tr} \left(\rho_{\rm red}^q \right) \, ,
\hspace{5mm} 
q>0 \, ,
\eeq
where $\rho_{\rm red}$ is the reduced density matrix computed earlier. 
In the limit $q\to1$ it reduces to the definition of entanglement entropy.
The eigenvalues of the reduced density matrix are known and for 
our theory it is given in eq. (\ref{eq:eigenRho}). This then becomes 
\beq
\label{eq:renEigen}
S_q =  \frac{1}{1-q} \log \sum_k p_k^q \, .
\eeq
For the eigenvalues $p_k$ given in eq. (\ref{eq:eigenRho})
one can perform the infinite summation easily as it is a 
geometric series. The expression for $S_q$ associated 
with each $SO(1,3)$ quantum number is given by,
\beq
\label{eq:SqRen}
S_q(p,\nu_{-}, \nu_{+}) = 
\frac{q}{1-q} \log (1 - \lvert\g_{-}\rvert^2)
+ \frac{q}{1-q}  \log (1 - \lvert\g_{+}\rvert^2)
- \frac{1}{1-q} \log \left(
1 - \lvert\g_{-} \g_{+}\rvert^{2q}
\right) \, .
\eeq
Then according as before we integrate $S_q(p,\nu_{-},\nu_{+})$ 
with the density of states for the $H^3$. This will give the piece of the 
R\'enyi entropy which contributes to long range
\beq
\label{eq:Sint_q}
S_{q, {\rm int}} \equiv c_6 
= \frac{1}{\pi} \int_0^\infty
{\rm d}p \, p^2 S_q(p, \nu_{-}, \nu_{+}) \, .
\eeq
Now various values of $q$ correspond to various interesting situations.
For $q\to0$, the R\'enyi entropy measures the dimension of the 
density matrix. It is also called Hartley entropy \cite{Headrick:2010zt}.
For $q\to1/2$ one gets a very simple expression for the $S_q$. 
\beq
\label{eq:Sq1/2Ren}
S_{\frac{1}{2}} (p,\nu_{-}, \nu_{+})
= \log \left[\frac{(1 - \lvert\g_{-}\rvert^2)(1 - \lvert\g_{+}\rvert^2)}
{\left(1 - \lvert\g_{-} \g_{+}\rvert \right)^2} \right] \, .
\eeq
In the case when $\nu_{-}=\nu_{+} = \nu$ then $S_{1/2} = 0$.
In figure \ref{fig:dsEERenM0} we plot the behaviour of 
R\'enyi entropy for various values of $q$ and massless non-local theory.

The particular value of q$=1/2$ has relation with entanglement negativity, which 
is a measure of the quantum entanglement between the two regions
\cite{Vidal:2002zz,Calabrese:2012ew,Calabrese:2012nk}. For our 
case it can be defined using the full density matrix 
$\rho = \ket{\Psi}\bra{\Psi}$, where $\ket{\Psi}$ is given in eq. (\ref{eq:psinew}). 
If the basis for the $R$ and $L$ region is denoted 
by $R^i$ and $L^i$, then the transposed matrix 
$(\rho_L)^T$ can be defined as $\bra{R^i L^j} (\rho_L)^T \ket{R^k L^m}
= \bra{R^i L^m} \rho \ket{R^k L^j}$. Using which one can 
express negativity as ${\cal E} \equiv \ln {\rm Tr} \lvert (\rho_L)^T \rvert$
where ${\rm Tr} \lvert (\rho_L)^T \rvert$ is the sum of the 
absolute value of the eigenvalues $(\rho_L)^T$. For pure state
\cite{Vidal:2002zz} then the negativity equals the value corresponding 
to $q=1/2$. This then immediately gives the finite interesting piece of 
the negativity as \cite{Maldacena:2012xp}
\beq
\label{eq:negFin}
{\cal E}_{\rm intr} = \frac{1}{\pi} \int_0^\infty
{\rm d} p p^2 S_{1/2} (p, \nu_{-}, \nu_{+}) \, ,
\eeq
where $S_{1/2} (p, \nu_{-}, \nu_{+})$ is given in eq. (\ref{eq:Sq1/2Ren}).

\section{Conclusions}
\label{conc}

In this paper we explore a non-local scalar-field theory. 
We start by showing that this non-locality can also be obtained 
starting from a local coupled system of two scalars by integrating 
out one of the scalar field. We further show that this coupled system 
can actually be expressed as a system of decoupled scalar fields
by a simple linear field transformation. In that sense the non-local 
system is equivalent to this. The aim of the paper is to seek an 
answer to question: how does the non-locality generated in 
infrared can affect or modify the low-energy physics of the system?

In cosmology we expect that there will exist entanglement beyond 
the Hubble radius, as in deSitter expanding space-time 
a particle pair which was initially created in causally connected region 
will eventually get separated and lie in causally disconnected regions
of deSitter space-time. We know that quantum corrections lead 
to infrared non-locality in theories. So if the system begets a 
non-locality which becomes important in infrared then such 
non-locality is expected to have a role to play in affecting the 
entanglement between two disconnected regions of the deSitter space-time. 
With this motivation we investigate the change in behaviour 
of the entanglement due to the presence of infrared non-locality. 

In local theories in flat space it is known that there is no long-range 
entanglement. However, in deSitter space it has been noticed 
in \cite{Maldacena:2012xp} that particle creation give rise to 
long-range entanglement. In local theories this feature is specific 
to deSitter space and doesn't have any flat space counterpart. 
In this paper we are interested in exploring the modification this
feature gets due to the presence of infrared non-locality
and ask the question whether non-locality can have 
any non-trivial effect on such long-range entanglement?

Following the methodology described in the paper \cite{Maldacena:2012xp},
we repeat the computation for the case of non-local massive scalar field 
theory. In the limit when non-locality strength goes to zero 
(\textit{i.e.} when theory becomes local)
we arrive at a local theory where the results are seen to be 
not in disagreement with the results obtained in \cite{Maldacena:2012xp}. 
In the case when we have non-locality in system then it adds 
an additional dimension into the problem. For massless theory, 
the presence of non-locality is seen to strongly affect the entanglement.
This effect is however not monotonic in nature. As $\lam^2$ is increased 
from zero to higher-values, it is noticed that the entanglement first 
increase, reaches a peak value and then decays. 
In the case when we change the sign of $\lam^2 \to -\lam^2$ then an oscillating 
behaviour is witnessed in the long-range entanglement. 
This situation has been depicted in the figure 
\ref{fig:dsEEm0}. Such oscillations is due to the presence of
tachyonic modes. This situation is gets more clear when one 
express the non-local system as a local system of decoupled scalars 
as expressed in eq. (\ref{eq:TransLagFT}). Then the massless non-local  case reduces 
to a very simple lagrangian of decoupled scalars with masses depending 
on $\lam$. Then the case with decaying entanglement refers to decoupled scalars
with complex masses, while the case of oscillating entanglement entropy 
refers to case of decoupled scalars with a tachyonic mass. 

In the case when we have a massive non-local theory then 
the behaviour of entanglement entropy is more 
complicated in nature. This is depicted in figure \ref{fig:dsEEmlam}. 
The qualitative features of this remains the same as in massless case 
although minute differences are noticed. 

We then computed R\'enyi entropy for our case. As the density matrix 
for the system was known, so it allowed the computation of R\'enyi 
easily following the definition. We then studied the behaviour of 
$S_q$ with respect to non-locality strength $\lam^2$ for various values of $q$. 
The qualitative features of $S_q$ are similar to the entanglement 
entropy: decaying for positive $\lam^2$ and oscillating for $-\lam^2$,
although a flattening of figure is witnessed in case of lower 
values of $q$. This is depicted in figure \ref{fig:dsEERenM0}.

The work presented here indicate the effect the presence 
of infrared non-locality will have on the entanglement between 
two disconnected regions of deSitter space-time. 
As infrared non-locality is an inevitable feature of well-defined 
quantum field theories, where it can naturally arise at low energies 
due to quantum corrections via renormalisation group running of 
parameters, therefore it means that infrared non-locality will have strong 
effect on the entanglement of disconnected regions. 
Hence, particle pair initially created in a causally connected 
region (inside Hubble radius) and later separated due to deSitter expansion 
(moved outside Hubble horizon) will be either strongly or weakly entangled 
depending on the nature of non-locality. 
This paper dealt with non-locality present in scalar field theory
which play an important role in cosmology in understanding 
both early and late-time Universe. However, similar 
analysis can also be extended to field theories of other spin 
particles. This work sheds light on the importance of non-locality 
in the long-range entanglement and the significant role they 
will play in the deSitter phase of the cosmology.

\bigskip
\centerline{\bf Acknowledgements} 
GN will like to thank Nirmalya Kajuri for useful discussion. 
GN is supported by ``Zhuoyue" (distinguished) Fellowship (ZYBH2018-03).
H. Q. Z. is supported by the National Natural Science Foundation of China (Grants
No. 11675140, No. 11705005, and No. 11875095).


\end{document}